 \documentclass[12pt]{article}
 \usepackage{epsfig,amsfonts,amssymb}
\usepackage[fleqn]{amsmath}
\usepackage{graphicx}
\usepackage{hhline}
\usepackage{cite}
\usepackage{upgreek}
\def\Xint#1{\mathchoice
    {\XXint\displaystyle\textstyle{#1}}%
    {\XXint\textstyle\scriptstyle{#1}}%
    {\XXint\scriptstyle\scriptscriptstyle{#1}}%
    {\XXint\scriptscriptstyle\scriptscriptstyle{#1}}%
    \!\int}
\def\XXint#1#2#3{{\setbox0=\hbox{$#1{#2#3}{\int}$}
    \vcenter{\hbox{$#2#3$}}\kern-.5\wd0}}

\def\dashint{\Xint-}
\begin{document}
\title{Energy spectrum of bound-spinons in the quantum Ising spin-chain ferromagnet}
\author{S.B. Rutkevich }
\date{December 6, 2007}
\maketitle
\begin{center}
\vspace{0.3cm}

{Institut f\"ur Theoretische Physik, Universit\"at M\"unster,\\
  Wilhelm-Klemm Str. 9, 48149 M\"unster, Germany, \\and \\
  Institute of Physics of Solids and Semiconductors, \\P. Brovka Str.  17,
 Minsk 220072, Belarus\\
 e-mail: rut@ifttp.bas-net.by}
\vspace{0.3cm}
\end{center}
\begin{abstract}
We study  the excitation energy 
spectrum in the $S=1/2$ ferromagnetic Ising spin chain with the easy axis $z$ in a magnetic 
field ${\mathbf h}=\{h_x,0,h_z\}$. According to
Wu and McCoy's scenario of weak confinement, the fermionic spinon excitations 
(kinks), being free at $h_z = 0$ in the ordered phase, are coupled into bosonic bound states
at arbitrary small $h_z > 0$. We  calculate the energy spectrum of
such excitations in the leading order in  small $h_z$, using different perturbative methods 
developed for the similar problem in the Ising field theory.
\end{abstract}

\section{Introduction} \label{SInt}
In recent years much progress has been achieved in the understanding of various 
aspects of the scaling limit of the  two-dimensional  
Ising model, which is known  as the  Ising Field Theory (IFT), for a review see \cite{Del04}. 
It is described by the Euclidean action
\begin{equation}
{\mathcal A}_{IFT}={\mathcal A}_{CFT}+\tau \int\varepsilon(x)\, d^2 x+h \int\sigma(x)\, d^2 x.
\label{AIFT}
\end{equation}
Here ${\mathcal A}_{IFT}$ corresponds to the conformal field theory in two dimensions, 
which is associated with the the Ising critical point. Fields $\varepsilon(x)$ (energy density)
and $\sigma(x)$ (spin density) are the only relevant operators in the theory, their  
scaling dimensions are $X_\epsilon=1$ and $X_\sigma=1/8$. Parameters $\tau$ and $h$ are proportional
to the deviation of the temperature $T$ and magnetic field $H$ in the lattice Ising model 
from their critical values: 
$\tau\sim T_c-T$ and $h\sim H $ at $T\to T_c$, $H\to 0$. In fact, only one dimensionless parameter 
$\eta=\tau/|h|^{8/15}$ determines the physics of IFT.

IFT is integrable along the directions $h=0$ and $\tau=0$. The former case $h=0$, $\tau\ne0$ 
corresponds to Onsager's solution \cite{Ons44}. Exact solution of IFT at $\tau=0$,  $h\ne0$
was found by A.B.  Zamolodchikov \cite{ZamH}. A thorough study of analytical properties of the free 
energy continued to complex values of the scaling parameter $\eta$ has been done by Fonseca 
and A.B.  Zamolodchikov\cite{FonZam2003} by means of a numerical technique, known as the truncated 
free-fermion space approach \cite{YuZam91}. Their analysis clarifies the relationship between the edge 
Yang-Lee singularity and the spinodal point.

For an analytical study of IFT for $h$ and $\tau$ close to the integrable directions, 
it is natural to exploit perturbation expansions. Form-factor perturbation theory developed by
Delfino, Mussardo and Simonetti \cite{Del96} has been applied to calculate the   variation of the 
particle mass spectrum and the decay widths of non-stable particle for small $\eta$, i.e. near the line
$\tau=0$  \cite{Del96,DGM06}. 

Perturbation theory for  IFT in the region close to the $h=0$ axis turns out to be rather nontrivial
in the low temperature phase $T<T_c$, as it was first shown by McCoy and Wu \cite{McCoy78}.
 At $h=0$ the particle sector of IFT contains one spinless fermion, 
which is interpreted in the ordered $T<T_c$ phase  as a kink interpolating between two degenerate vacua. 
Application of magnetic field removes the degeneracy and induces a long-range  interaction between 
kinks, which leads to their confinement into pairs.  This means, that even at  small field $h$ an 
isolated kink gains an infinite energy, and the bounded kink-antikink pairs become the only 
single-particle excitations in the model. Their dispersion law $E_n(P)$ has the relativistic form: 
\begin{equation}
E_n(P)=\sqrt{P^2+M_n^2},  \label{Ld}
\end{equation}
being completely determined by the masses $M_n$. 
This follows from the Lorentz invariance of IFT, which in turn is the result
of the rotation invariance of the two-dimensional Ising model in the scaling limit.

At small $h$ the mass spectrum $M_n$  of kink-antikink pairs 
becomes dense in the segment $[2m,\infty)$. Two  asymptotic expansions describe  $M_n$ 
at $h\to0$ in different regions of this segment. Near the  edge point $2m$ (i.e. for
fixed $n$ at $h\to0$)   one can use the {\it low energy} expansion in fractional powers 
of the magnetic field
\begin{eqnarray} \label{low}
\frac{M_n^2}{4 m^2}=1+\sum_{k=2}^\infty \mu_{k} \zeta^{k/3},
\end{eqnarray} 
where  $m=2\pi \tau$ is the mass of a free fermion (kink) at $h=0$, 
 $\bar{\sigma}$ is the spontaneous
magnetization at $h=0$, and $\zeta=2 h \bar{\sigma}/m^2$. Several initial coefficients $\mu_{k}$  are known explicitly.
The coefficient of the leading term has been found by McCoy and Wu \cite{McCoy78} in the form $\mu_{2}=z_n$, 
where $-z_n$ denotes the zeroes of the Airy function, ${\rm Ai}(-z_n)=0$.  Correction terms in (\ref{low})
 up to $k=8$ have been calculated by Fonseca and A.B. Zamolodchikov, see \cite{FonZam2003,FZ06}.

On the other hand, for $n\gg1$ and  $h\to0$  the {\it semiclassical} expansion in integer powers 
of $h$ can be applied:
\begin{equation} \label{wkbe}
\frac{M_n^2}{4 m^2} =  [1+a_2 \,\zeta^2+O(\zeta^3)]\,\cosh \vartheta_n,
\end{equation}
where \cite{FZWard03}  $a_2=0.071010809\ldots$, and the numbers $\vartheta_n$ denote the solutions of  equation 
\begin{equation} \label{bsccc}
\sinh 2 \vartheta_n -2 \vartheta_n=2 \pi \zeta\, (n-1/4)-\zeta^2 \bar{S}_1(\vartheta_n)-O(\zeta^3),
\end{equation}
with 
\begin{equation} \label{wkbd}
{\bar S}_1 (\vartheta) =
-{1\over{\sinh 2\vartheta}}\,\bigg[{5\over
    24}\,{1\over{\sinh^2\vartheta}} + {1\over 4}\,{1\over{\cosh^2\vartheta}}
  - {1\over 12} - {1\over 6}\,\sinh^2\vartheta\bigg].
\end{equation}
The leading  terms of order $\zeta$ in (\ref{wkbe}), (\ref{bsccc}) were  determined  by Rutkevich \cite{Rut05}, 
and independently by Fonseca and A.B. Zamolodchikov \cite{FZ06}. The second-order term (\ref{wkbd}) 
in (\ref{bsccc}) has been found by Fonseca and A.B. Zamolodchikov \cite{FZ06}. 

IFT viewed as a particle theory gives us a nice 
model  of quark confinement. It is worth to note, that there are a lot of similarities
between confinement in IFT and in  't~Hooft's model for  the two-dimensional
multicolor QCD \cite{Hooft74}, see the discussion in \cite{FZ06}. 
On the other hand, IFT can be treated as  the continuous model of the one-dimensional quantum ferro-
(or antiferro-) magnet. 
The continuous approximation applies to one-dimensional ferromagnets only in the critical region 
near the quantum phase transition point  \cite{Sach99}. Beyond the critical region the discreteness of the  lattice 
become important. 

The purpose of the present  paper is to discuss, how the picture of 
confinement outlined above is  modified by the discrete lattice effects beyond the critical region. 
Preliminary results in this directions have been obtained in \cite{Rut07}.
Our work is motivated by 
recent experimental observation of confinement of topological excitations in spin-$1/2$ chain antiferromagnet 
reported by Kenzelmann {\it et al.} \cite{Kenz05}.
 Accordingly, we shall hold to the magnetic terminology: the terms ``spinon'', ``domain wall'' or "kink" will be used 
for  topological excitations, and  their  bound-states will be called ``bound-spinons'' \cite{Kenz05}.
The problem will be considered within the simplest 
appropriate one-dimensional  model of ferromagnetism:  
$S=1/2$ Ising spin chain in a ``skew'' magnetic field  \cite{Fog78,Tsv04}, which has components 
both normal and parallel to the easy magnetization axis $z$, ${\mathbf h}=\{h_x,0,h_z\}$.  This model has an 
exact solution in 
the case of a purely transverse magnetic field ${\mathbf h}=\{h_x,0,0\}$ and reduces to IFT 
in the critical region near the quantum phase 
transition point \cite{Sach99}. We  focus our attention on the dispersion law of bound-spinons 
in the limit of small longitudinal 
magnetic field $h_z\to0$, and extend the results (\ref{low}),  (\ref{wkbe}) 
to this model beyond the critical region\footnote{Of course, the relativistic functional form (\ref{Ld})
does not hold  for the bound-spinon spectrum  in the discrete  spin chain model apart from the critical region.}.
It turns out, that  the bound-spinon energy spectrum in the discrete spin chain exhibits a rather rich   
small-$h_z$ asymptotical  behavior. We describe eight different 
 regimes for the bound-spinon energy spectrum, which are realized in the noncritical 
Ising spin chain in the limit $h_z\to0$, instead of two regimes (\ref{low}), (\ref{wkbe}) known  in IFT.
 
The rest of the paper is organized as follows. In Section \ref{SModel} the spin-$1/2$ Ising chain model 
in a skew magnetic field is described. We study the bound-spinon spectrum in this model 
by means of two different procedures, which are shown to be effective  in the IFT \cite{Rut05,FZ06}. 
In Section \ref{SBound} we apply a heuristic approach based on the Bohr-Sommerfeld quantization rule.
Results obtained in  Section \ref{SBound}  are confirmed and extended in more systematic theory in subsequent 
sections. In Section \ref{S2FA} we derive a singular integral equation, which determines the energy spectrum 
of  bound-spinons in the two-fermion approximation, and  generalizes to the  discrete spin chain model 
the IFT Bethe-Salpeter equation \cite{FonZam2003,FZ06}.
Section \ref{Sec5} contains some technical results relating to the singular integral equation (\ref{ieq}).
  In Section \ref{Toy} we derive  exact solutions of two simplified versions of this equation, 
and give explicit expressions for the corresponding bound-spinon energy spectra. 
In Section \ref{Sec7} we return to the original integral equation (\ref{ieq}), and develop for it a weak coupling 
perturbative expansion based on the results of Section \ref{Sec5}. The lowest order of this expansion 
leads to  eight different  small-$h_z$ asymptotics for the bound-spinon dispersion law, which can be used in different
regions of the bound spinon quasimomentum and energy. Obtained results are discussed in Section \ref{Dis}.
\section{Ising model in skew magnetic field} \label{SModel}
 The quantum Ising spin-$1/2$ chain model is defined by the Hamiltonian:
\begin{equation}
\label{Ham}
{\mathcal H} =-\sum_{j} (\sigma_j^z \,\sigma_{j+1}^z+ h_x \sigma_j^x +h_z \sigma_j^z ).
\end{equation}
Here $\sigma ^{x,z}$  are the Pauli matrices, $j$ enumerates the chain sites.  The number of
the chain sites is put to infinity in the thermodynamical limit. The applied ``skew'' magnetic
field ${\mathbf h}$ has both components parallel and normal to the magnetic easy axis $z$.
The Hamiltonian is normalized to the ferromagnetic nearest neighbor coupling constant.
\begin{figure}[ht]
\centering
\includegraphics[width=.95 \linewidth]{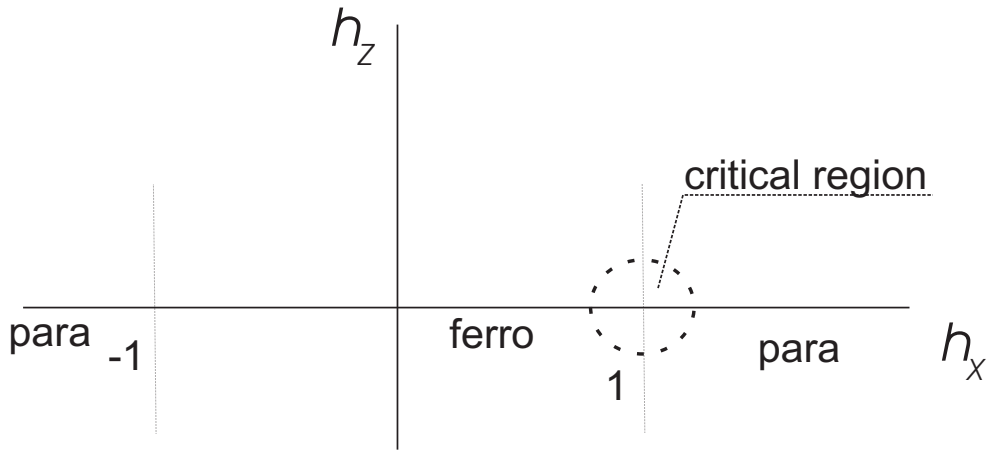} 
\caption{\label{Diag}Phase diagram of the Ising spin chain at $T=0$.}
\end{figure}

The phase diagram of  model (\ref{Ham}) at zero temperature is shown in Fig. \ref{Diag}. 
Model (\ref{Ham}) is not integrable at generic nonzero
$h_x,\;h_z$. However,
in the line $h_z=0$ it has  exact solution first found by Pikin and Tsukernik \cite{Tsuk}.
At $h_x=\pm 1$ lie the quantum phase transition points, which separate ferromagnetic
and paramagnetic phases  \cite{Sach99}. In the critical regions near these points the model is 
equivalent to IFT.
Two ferromagnetic ground states $\mid 0_{+}\rangle $ and $\mid 0_{-}\rangle$
coexist in the ferromagnetic phase $-1<h_{x}<1$. These ground states have the same energies 
$E_+ (h_z=0)=E_- (h_z=0)$, but the opposite signs of
the spontaneous magnetization $\langle 0_{\pm }\mid \sigma _{j}^{z}$ $\mid
0_{\pm }\rangle =\pm $ $\bar{\sigma}$, where $\bar{\sigma}=(1-h_{x}^{2})^{1/8}$.

In the integrable line $h_z=0$ the model Hamiltonian (\ref{Ham}) can be reduced to 
free fermions by use of the Jordan-Wigner transformation \cite{Tsuk,Jimb80}:
\begin{equation}
{\cal H}_{0}\equiv {\cal H}_{ch}\mid_{h_{z=0}}=\int_{-\pi }^{\pi }\frac{\rm{d}%
\theta }{2\pi }\ \omega (\theta )\,a^{\dagger }(\theta )\,a (\theta )+%
{\rm const},  \label{Ht}
\end{equation}
where $\theta $ is the quasi-momentum, fermionic operators $a ^{\dagger
}(\theta )$,  $a(\theta )$ satisfy the canonical anticommutation
relations
\begin{eqnarray}
\left\{ a (\theta )\ ,a (\theta ^{\prime })\right\}  &=&\left\{ a   \nonumber
^{\dagger }(\theta )\ ,a^{\dagger }(\theta ^{\prime })\right\} =0, \\
\left\{ a ^{\dagger }(\theta )\ ,a (\theta ^{\prime })\right\}
&=&2\pi \delta (\theta -\theta ^{\prime }),  \nonumber
\end{eqnarray}
and the free-fermion dispersion law is given by
\begin{equation}
\label{fdis}
\omega (\theta )=2\left[ (1-h_{x})^{2}+4h_{x}\sin ^{2}\frac{\theta }{2}%
\right] ^{1/2}.
\end{equation}
Operators $\sigma_j^x$ and $\sigma_j^z \sigma_{j+1}^z $ are  bilinear in fermionic fields.
Operator $\sigma_j^z$ can be written as a normally ordered exponential of a bilinear form of fermionic
operators \cite{Jimb80}. It can be also completely characterized by the  form-factors:  
\begin{equation*} 
\langle \theta_1,\ldots, \theta_K|\sigma_0^z|\theta_1'\ldots,\theta_N'\rangle =
\langle 0|a(\theta_1),\ldots, a(\theta_K) \,
\sigma_0^z\,  a^{\dagger}(\theta_1'),\ldots,a^{\dagger}(\theta_N')|0\rangle,
\end{equation*}
where $(K+N)$ takes even values. The elementary form-factors are given by:
\begin{eqnarray} \label{elform}
\bar{\sigma}^{-1}\langle\theta_1,\theta_2|\sigma_0^z|0\rangle=F(\theta_1,\theta_2|)=
\frac{1}{1-\exp[i(\theta_1+\theta_2)]}\,\frac{\omega(\theta_1)-\omega(\theta_2)}
{\sqrt{\omega(\theta_1)\omega(\theta_2)}},\\
\bar{\sigma}^{-1}\langle 0|\sigma_0^z| \theta_1,\theta_2\rangle=F(|\theta_1,\theta_2)=
\frac{1}{\exp[-i(\theta_1+\theta_2)]-1}\,\frac{\omega(\theta_1)-\omega(\theta_2)}
{\sqrt{\omega(\theta_1)\omega(\theta_2)}}, \\
\bar{\sigma}^{-1}\langle \theta|\sigma_0^z| \theta'\rangle=F(\theta|\theta')=
\frac{1}{1-\exp[i(\theta-\theta')]}\,\frac{\omega(\theta)+\omega(\theta')}
{\sqrt{\omega(\theta)\omega(\theta')}}.
\end{eqnarray}
All other form-factors can be obtained from the elementary ones by application of the Wick rule. 
For example, the form-factor with two fermions in the initial and final states can 
be written as:
\begin{equation}  \label{form2}
\frac{\langle \theta_1,\theta_2|\sigma_0^z|\theta_1'\theta_2'\rangle}{\bar{\sigma}} =
F(\theta_1,\theta_2|)F(|\theta_1',\theta_2')+F(\theta_1|\theta_2')F(\theta_2|\theta_1')
-F(\theta_1|\theta_1')F(\theta_2|\theta_1').
\end{equation}
\section{Bound states of two domain walls} \label{SBound}
In the ferromagnetic phase $|h_x|<1$, the free fermions in (\ref{Ham}) represent the domain walls, 
which separate regions with different orientations of the magnetization. A small longitudinal 
magnetic field $h_z>0$, which breaks the  $\mathbb{Z}_2$-symmetry,  evidently provides the 
long range attraction force $\chi=2  h_z \bar{\sigma}$ between the 
two neighboring domain walls. Due to this confining force, 
all domain walls become coupled into pairs which we shall call bound-spinons, following  \cite{Kenz05}.
If $h_z$  is small, the weak confinement regime is realized.  
The energy spectrum of bound-spinons in this regime can be understood to much 
extent in the following heuristic approach, first developed for the IFT 
\cite{McCoy78,FonZam2003,FZ06,Rut05} . 

Consider two interacting  fermions moving in a line   as a classical system with the Hamiltonian
\begin{equation}
\label{Hcl}
\mathcal{H}(\theta_1,\theta_2,x_1,x_2)= \omega(\theta_1)+ \omega(\theta_2)+ \chi\, |x_2-x_1|
\end{equation}
Here the fermion coordinates   $x_1, \, x_2 $ are the real numbers.  
Variables $\theta_1, \,\theta_2$ are the canonical momenta corresponding to the coordinates $x_1, \, x_2 $, 
and the fermion  kinetic energy $\omega(\theta)$ is given by (\ref{fdis}).
After the canonical transformation
\begin{eqnarray} \nonumber
X&=&\frac{x_1+x_2}{2},\;\; x=x_2-x_1,\\
\Theta&=&\theta_1+\theta_2,\,\theta=\frac{\theta_2-\theta_1}{2}, \label{th12}
\end{eqnarray}
the Hamiltonian (\ref{Hcl}) takes the form
\begin{equation}
\label{Hcl1}
\mathcal{H}(\theta,x;\Theta)= \epsilon(\theta;\Theta)+ \chi\, |x|,
\end{equation}
where  
\begin{equation} \label{eps}
\epsilon(\theta;\Theta)= \omega(\theta+\Theta/2)+ \omega(\theta-\Theta/2). 
\end{equation}
The total energy-momentum conservation laws 
read as:
\begin{eqnarray}
&&\epsilon(\theta(t);\Theta)+\chi\, |x(t)|=E={\rm Const},\label{cons}\\
&&\Theta (t)={\rm Const}.  \nonumber         
\end{eqnarray}
\begin{figure}[h!]
     \leavevmode
\centering
\includegraphics[width=.8 \linewidth, angle=00]{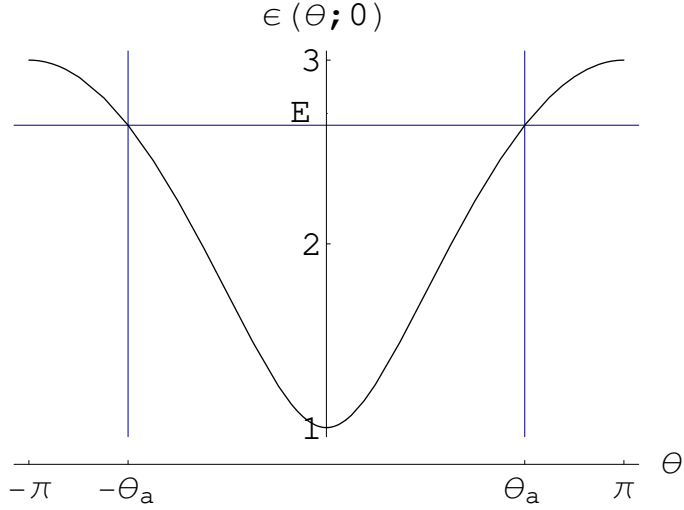} 
\caption{\label{F2}Function $\epsilon(\theta;\Theta)$ given by (\ref{eps}), (\ref{fdis})  for $\Theta=0$ and  $h_x=0.5$.
For energy $E$ classically allowed region is $[-\theta_a,\theta_a]$.}
\end{figure}
The canonical equations of motion are: 
\begin{eqnarray}
&&\dot{X}(t)=\frac{\partial\epsilon(\theta;\Theta)}{\partial \Theta}, \label{emo}\\
&&\dot{x}(t)=\frac{\partial \epsilon(\theta;\Theta)}{\partial \theta},\label{emox}\\
&&\dot{\theta}(t)=-\chi \, {\rm sign}[ x(t)]. \label{th}
\end{eqnarray}
For a given  value of the conserved total momentum $\Theta$, Hamiltonian (\ref{Hcl1}) describes the  relative 
motion of two fermions.  This motion becomes especially simple in the
$\theta$-space.

Fig. \ref{F2} shoes  the $\theta$-dependence of the ``kinetic energy''  $\epsilon(\theta;\Theta)$
for $h_x=0.5$ and $\Theta=0$.  
If $\theta=\theta_a$ and $x=+0$ at $t=0$, the 
total energy takes the value $E=\epsilon(\theta_a,\Theta)$. Due to (\ref{th}), the
momentum $\theta$  of the relative motion  will linearly decreases in time
\begin{equation}
\theta(t)= \theta_a-\chi\,t  \label{th(t)}
\end{equation} 
until the moment $t=t_1=2\theta_a/\chi$, when $\theta(t)$  reaches the value $-\theta_a$.
 The time variation of the space coordinate $x(t)$ for $0<t<t_1$ can be read
from (\ref{cons}):
\begin{equation}
x(t)= \frac{E-\epsilon[\theta(t);\Theta]}{\chi}. \label{x(t)}
\end{equation}

Then $x(t)$ becomes negative, and in the interval $t_1<t<2 \,t_1$ canonical coordinates
vary as
\begin{eqnarray*}
\theta(t)= -\theta_a+ (t-t_1)\,\chi,\\
x(t)= -\frac{E-\epsilon[\theta(t);\Theta]}{\chi}.
\end{eqnarray*}
Then the phase trajectories periodically repeat in time with the period $2\,t_1$.
The semiclassical energy levels can be obtained from the Bohr-Sommerfeld quantization condition:
\begin{equation}
\oint dx \,\theta = 2 \pi (\nu+1/2), \label{BBS}
\end{equation}
where integration is taken along the closed periodic phase path in the $(x,\theta)$-plane
over one cycle of motion, $0<t<2\,t_1$. Since the two interacting particles are fermions, 
only odd values are allowed
for the integer $\nu$ due to the Pauli principle: 
\begin{equation}
\label{odd}
\nu=2n-1, \quad n=1,2,\ldots \,\,.
\end{equation}
The right-hand side of (\ref{BBS}) can be transformed as
\begin{eqnarray} \label{ider}
\oint dx \,\theta=-2 \int_{-\theta_a}^{\theta_a} d\theta \,\theta\, \frac{d x(\theta)}{d\theta }
=\frac{2}{\chi}  \int_{-\theta_a}^{\theta_a} d\theta \,\theta\; \dot{x}(t)=\\
\frac{2}{\chi} \int_{-\theta_a}^{\theta_a} d\theta \,\theta\;
\frac{\partial \epsilon(\theta;\Theta)}{\partial \theta}=\frac{2}{\chi}
\Bigg(2 E\,\theta_a-\int_{-\theta_a}^{\theta_a} d\theta\,\epsilon(\theta;\Theta)\Bigg).    \nonumber
\end{eqnarray}
Here in the second equality we have taken into account the linear dependence (\ref{th(t)}) between 
$\theta$ and $t$ for $0<t<t_1$, and in the third equality we have used  equation of motion
(\ref{emox}). Combining (\ref{BBS}), (\ref{odd}), and (\ref{ider}) we find the semiclassical
energy spectrum $E_n(\Theta)$ of bound states in the system of two fermions (\ref{Hcl}):
\begin{equation}
2 E_n(\Theta)\, \theta_a-\int_{-\theta_a}^{\theta_a} d\theta \, \epsilon (\theta,\Theta)  =
2 \pi\, \chi \,(n-1/4)  , \qquad n=1,2,\ldots, \label{qrule}
\end{equation}
where $\theta_a\in [0,\pi]$ is the solution of equation
\begin{equation}
\epsilon (\theta_a;\Theta)=E_n(\Theta). \label{thet}
\end{equation}

\begin{figure}[h!]
     \leavevmode
\centering
\includegraphics[width=.8 \linewidth, angle=00]{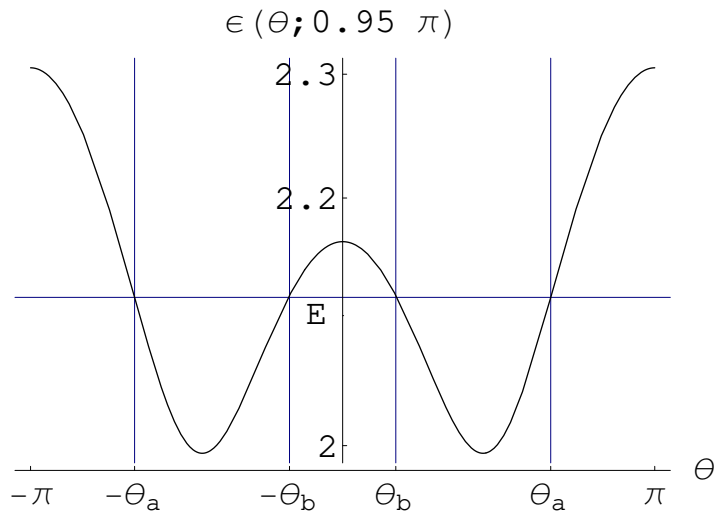} 
\caption{\label{F3}Function $\epsilon(\theta;\Theta)$  determined by  (\ref{eps}) (\ref{fdis}) 
for $\Theta=0.95 \pi$ and  $h_x=0.5$.
For energy $E$ classically allowed regions are $[-\theta_a,-\theta_b]$ and $[\theta_b,\theta_a]$. }
\end{figure}

In the above treatment  we implied that function $\epsilon(\theta;\Theta)$ monotonically
increases with $\theta$ in the interval $0<\theta<\pi$, providing that equation (\ref{thet}) has a single
solution there for  $E\in [\epsilon(0;\Theta),\epsilon(\pi;\Theta)]$.  
This is true for small enough values of the total 
momentum $\Theta$: 
\begin{equation*} 
0<\Theta<\Theta_m, \quad \Theta_m=2 \arccos h_x .
\end{equation*}
However, for  $\Theta>\Theta_m$ the second derivative $\partial^2 \epsilon(\theta;\Theta)/\partial \theta^2$ at 
$\theta=0$
becomes negative, see Fig. \ref{F3}. 
Though for high enough energies  $\epsilon(0,\Theta)<E<\epsilon(\pi,\Theta)$  we can still 
use  formula (\ref{qrule}), for smaller energies
 $E<\epsilon(0,\Theta)$, equation  (\ref{cons}) has  
 two solutions $\theta_a, \theta_b$ in the 
interval $0<\theta<\pi$. In the latter case the classical phase trajectory $\theta(t)$ moves back 
and forth in one of the the classically allowed region, say in the interval  $[\theta_b,\theta_a]$. 
Equations (\ref{th(t)}), (\ref{x(t)}) remain still valid, 
 with $t_1=(\theta_a-\theta_b)/\chi$, however. 
Semiclassical energy levels are determined by condition (\ref{BBS}), where the integration path
is located now in the interval  $[\theta_b, \theta_a]$. Since this interval is not invariant under 
reflection $\theta\to-\theta$,   the Pauli principle does not impose  any restrictions 
on the integers  $\nu=0,1,2,\ldots$ in (\ref{BBS}). Therefore, instead of (\ref{qrule}), we get 
for $\Theta_m<\Theta<\pi$ and $E<\epsilon(0,\Theta)$: 
\begin{equation}
 E_n(\Theta)\, (\theta_a-\theta_b)-\int_{\theta_b}^{\theta_a} d\theta \, \epsilon (\theta,\Theta)  =
 \pi\, \chi \,(n-1/2)  , \qquad n=1,2,\ldots, \label{qrule1}
\end{equation}
where $n=\nu+1$, and $\theta_a,\theta_b \in [0,\pi]$ are the solutions of equation
\begin{equation}
\epsilon (\theta_{a,b}\,,\Theta)=E_n(\Theta), 
\end{equation}
shown in Fig. \ref{F3}.

Semiclassical formulas (\ref{qrule}), (\ref{qrule1}) can be used,  if  $n$ is large,  $n\gg1$, and the 
energy $E_n(\Theta)$ does not approach  one of the critical values $E_c(\Theta)$ of the function  
$\epsilon(\theta,\Theta)$: $E_c(\Theta)=\epsilon(\theta_c,\Theta), \quad \partial\epsilon(\theta,\Theta)
/ \partial  \theta  |_{\theta=\theta_c}=0$.
On the other hand, the low-energy part of the spectrum for small $h_z\to 0$ and $0<\Theta<\Theta_m$ 
can be determined  by use of the ``nonrelativistic approximation'' developed for the IFT
\cite{McCoy78,FonZam2003,FZ06}. One should expand the ``kinetic energy'' in $\theta$ near the 
origin $\epsilon(\theta,\Theta)\approx \varepsilon_0(\Theta)+ \varepsilon_2(\Theta)\theta^2/2 $ reducing the
problem to the Schr\"odinger equation with the Hamiltonian 
$ \varepsilon_0(\Theta)-\frac{1}{2}\varepsilon_2(\Theta)\partial_x^2+
\chi|x|$. Only odd wave eigenfunctions $\phi(x)$ should be taken into account due to the Pauli principle, 
and the energy levels take the form
\begin{equation} \label{ei}
E_n(\Theta)\approx \varepsilon_0(\Theta)+\chi^{2/3}[\varepsilon_2(\Theta)/2]^{1/3}\,z_n,
\end{equation}
where 
\begin{equation}
\varepsilon_0(\Theta)=\epsilon(0,\Theta), \quad 
\varepsilon_2(\Theta)=\partial^2 \epsilon(\theta,\Theta)/\partial \theta^2|_{\theta=0},  
\end{equation}
and $-z_n, \; n=1,2,\ldots$ are the zeros of the Airy function ${\rm Ai}(-z_n)=0$.
\section{Two-fermion approximation in the Ising chain  model} \label{S2FA}
The bound-spinon energy spectrum can be formally defined as the solution of the eigenvalue
problem
\begin{equation}\label{MP}
({\mathcal H} -E_{\rm vac})\mid \Phi_n(\Theta)\rangle = E_n(\Theta) \mid \Phi_n(\Theta)\rangle,
\end{equation}
where  ${\mathcal H}$ is the Ising chain Hamiltonian (\ref{Ham}),  $E_{\rm vac}$ is the ground state energy, 
and $|\Phi_n(\Theta)\rangle$ is the bound-spinon eigenvector with quasimomentum $\Theta$:
$\hat{T}_1 |\Phi_n(\Theta)\rangle= \exp(i \Theta)\, |\Phi_n(\Theta)\rangle$, with $\hat{T}_1$ being the lattice site 
translation operator. 
In the small-$h_z$ limit this problem can be studied within the two-fermion approximation, which
turns out to be effective in IFT \cite{FonZam2003,FZ06,Rut05}. In this approach one considers 
instead of the  the 
exact eigenvalue problem (\ref{ei}) its projection to the two-fermion subspace $\mathcal F_2 $ of the Fock space 
\begin{equation}
\mathcal P_2 \,\mathcal H  \, \mathcal P_2 \mid 
\Phi_n(\Theta)\rangle = E_n(\Theta)\,\mathcal P_2 \mid \Phi_n(\Theta)\rangle,  \label{eig2}
\end{equation}
where $\mathcal P_2$ is the orthogonal projector onto $\mathcal F_2 $. 
So, the bound-spinon state is approximated by a two-fermion state, neglecting the four-fermion, six fermion, and higher
multi-fermion contributions. In the momentum representation (\ref{eig2}) reads as
\begin{eqnarray} \label{bs1}
[\omega(\theta_1)+\omega(\theta_1)-E_n(\Theta)]\phi_n(\theta_1,\theta_2)=\frac{h_z}{2}
\int\!\!\!\int_{-\pi}^\pi \frac{d\theta_1'\,d\theta_2'}{2 \pi} \phi_n(\theta_1',\theta_2') \cdot\\
\sum_k\delta(\theta_1'+\theta_2'+2 \pi k-\Theta)
\langle \theta_2,\theta_1|\sigma_0^z|\theta_1'\theta_2'\rangle,   \nonumber
\end{eqnarray}
where 
\begin{eqnarray} \nonumber
\phi_n(\theta_1,\theta_2)= \langle\theta_2,\theta_1|\mathcal P_2|\Phi_n(\Theta)\rangle, \\ 
\phi_n(\theta_2,\theta_1)=-\phi_n(\theta_1,\theta_2), \label{sq}\\
\theta_1+\theta_2=\Theta+2 \pi k, \quad k=0,\pm 1,\pm 2 \ldots. \nonumber
\end{eqnarray}
Taking into account (\ref{form2}) and  (\ref{sq}), we can replace the two-fermion form-factor in the integrand, as  
\begin{eqnarray*}
\langle \theta_2,\theta_1|\sigma_0^z|\theta_1'\theta_2'\rangle \to 4\,\bar{\sigma}\, 
{\mathcal G}(\theta_2,\theta_1|\theta_1',\theta_2'),\\
 {\mathcal G}(\theta_2,\theta_1|\theta_1',\theta_2')=\frac{1}{4}\,[F(\theta_2,\theta_1|)F(|\theta_1',\theta_2')+
2\,F(\theta_1|\theta_1')F(\theta_2|\theta_2')],
\end{eqnarray*}
where the elementary form-factors $F(\ldots)$ are given by (\ref{elform}).
Since the bound-spinon quasimomentum $\Theta$ is fixed, one can rewrite (\ref{bs1})  in the 
variables $\theta=\theta_2-\theta_1$ and $\theta'=\theta_2'-\theta_1'$ :
\begin{eqnarray} \label{bs2}
[\epsilon(\theta;\Theta)-E_n(\Theta)]\,\phi_{n,\Theta}(\theta)=\chi\,
\dashint_{-\pi}^\pi \frac{d\theta'}{2 \pi} G_\Theta(\theta,\theta')\, \phi_{n,\Theta}(\theta'),
\end{eqnarray}
where
$\epsilon(\theta;\Theta)$ is given by (\ref{eps}), $\chi=2 h_z \bar{\sigma}$, and 
\begin{equation} \label{G_Th}
G_\Theta(\theta,\theta')={\mathcal G}\Bigg(\frac{\Theta}{2}+\theta,\frac{\Theta}{2}-\theta\Bigg|
\frac{\Theta}{2}-\theta',\frac{\Theta}{2}+\theta',\Bigg)
\end{equation}
The kernel has the second order pole singularity at $\theta=\theta'$:
\begin{equation} \label{Gre}
G_\Theta(\theta,\theta')=-\frac{2}{\{\exp[i(\theta- \theta')/2]-\exp[-i(\theta- \theta')/2]\}^2}+
G_\Theta^{(reg)}(\theta,\theta'),
\end{equation}
where $G_\Theta^{(reg)}(\theta,\theta')$ is regular at real $\theta$ and $\theta'$.

The function $\phi_{n,\Theta}(\theta)$ in equation (\ref{bs2}) is odd and $2\pi$-periodic in $\theta$, 
and the integral 
in the right-hand side is understood in the sense of the Cauchy principal value. Equation (\ref{bs2}) 
gives the discrete-lattice
version of the IFT Bethe-Salpeter equation in a generic momentum frame, see (3.11) in \cite{FZ06}.
It is instructive to rewrite  (\ref{bs2}) in the coordinate representation:
\begin{eqnarray} \label{bsco} 
\sum_{j'}K(j-j')\,\phi_{n,\Theta}(j')-E_n(\Theta)\,\phi_{n,\Theta}(j)= \\\nonumber
-\chi\,|j|\, \phi_{n,\Theta}(j)
+\chi\,\sum_{j'}{\mathcal U}(j,j')\,\phi_{n,\Theta}(j'),
\end{eqnarray}
where $K(j)$ and $\phi_{n,\Theta}(j)$ are the Fourier coefficients of the $2\pi$-periodic functions
$\epsilon(\theta;\Theta)$ and $\phi_{n,\Theta}(\theta)$, respectively, $j,j'=0,\pm 1,\ldots$.

The first and the second terms in the right-hand side of (\ref{bsco}) come, respectively, from the 
first  and the second terms in (\ref{Gre}). The first term  describes the long-range attraction 
between fermions (domain walls), while the second, being  responsible for the short-range interaction, 
 vanishes  exponentially in  $|j|$ and $|j'|$ on  the correlation length scale.

In this paper we shall focus our attention on the  perturbative solution of the singular 
integral equation (\ref{bs2}) in the leading order in the small parameter $\chi$. As we know from 
the IFT, the short-range interaction does not effect  the fermion bound-state energy 
in the leading order in $h_z$, contributing only to the higher order corrections \cite{Rut05,FZ06}.
Accordingly, in what follows
we shall drop the second term in the right-hand side of (\ref{Gre}), which does not contribute
to the bound-spinon energy $E_n(\Theta)$ in the leading order in $\chi$. Then, equation (\ref{bs2})
 takes the form in variables $z=\exp(i\theta)$, $z'=\exp(i\theta')$:
\begin{equation} \label{ieqA}
[\epsilon(z;\Theta)-E(\Theta)]\,\phi(z)=-  \chi\,z \dashint_{S_1}\frac{dz'}{\pi i}\, \frac{\phi(z')}{(z'-z)^2},
\end{equation}
where
\begin{equation} \label{eis}
\epsilon(z;\Theta)=2\sqrt{h_x}\Bigg\{\Bigg[h_x+\frac{1}{h_x}- zv -\frac{1}{zv}\Bigg]^{1/2}+
\Bigg[h_x+\frac{1}{h_x}- \frac{z}{v} -\frac{v}{z}\Bigg]^{1/2}\Bigg\},
\end{equation}
and  $v=\exp(i\Theta/2)$.
We use notation $S_r$ for  the  circle of the radius $r$ centered in the origin and 
passed in the counter-clockwise direction, and skip indices at $E_n(\Theta)$ and $\phi_{n,\Theta}(z)$. 
\section{Singular integral equation in the unit circle} \label{Sec5}
In the present Section we describe  some  general  properties of equation (\ref{ieqA}), which will be used in Section 
\ref{Toy} and \ref{Sec7}. 

Consider a more general singular integral equation\footnote{
On the general theory of singular integral equations see \cite{Mus}.}
\begin{equation} \label{ieq}
[\epsilon(z)-\lambda]\,\phi(z)=-  \mu\,z \dashint_{S_1}\frac{dz'}{\pi i}\, \frac{\phi(z')}{(z'-z)^2}.
\end{equation}
Functions $\epsilon(z)$ and $\phi(z)$  are   supposed to be analytical in a narrow ring $\Gamma$, 
$\Gamma=\{z\in\Gamma\,|\, 1-\delta<|z|<1+\delta\}$, with small positive  $\delta$.
The symmetry properties $\epsilon(1/z)=\epsilon(z)$ and  $\phi(1/z)=-\phi(z)$
are also  required. 

Let us define two functions $g_+(z)$ and $g_-(z)$: 
\begin{equation} \label{gpm}
g_\pm (z)= \oint_{S_1}\frac{dz'}{2\pi i}\, \frac{\phi(z')}{(z'-z)},
\end{equation}
where  $g_+(z)$ is defined at $|z|<1$, and  $g_-(z)$ is defined in the region
$|z|>1$. 
The evident properties of these function are:
\begin{enumerate}
\item  $g_+(z)$ and $g_-(z)$  are  analytical at $|z|< 1$ and at $|z|> 1$,  respectively. \label{anal}
\item  $g_+(z)$ and $g_-(z)$ can be continued to the unit circle $S_1$, where they are continuous together
with their derivatives.   
\item $g_+(1/z)=g_-(z)$.
\item $g_+(0)=g_-(\infty)=0.$
\item $g_+(z)-g_-(z)=\phi(z)$ for $|z|=1$. \label{i5}
\item $\partial_z g_+ (z)+\partial_z g_-(z)=(\pi i)^{-1} \dashint_{S_1}dz'\, \phi(z')\,(z'-z)^{-2}$  for $|z|=1$.
\label{i6}
\end{enumerate}

Combining properties \ref{i5}, \ref{i6} with (\ref{ieq}),  one finds for $|z|=1$
\begin{equation} \label{U}
\{[\epsilon(z)-\lambda]+\mu\,z\,\partial_z\}\,g_+(z)=U(z)=\{[\epsilon(z)-\lambda]-\mu\,z\,\partial_z\}\,g_-(z).  
\end{equation}
The function $U(z)$ defined this way is invariant with respect to the inversion
\begin{equation} \label{invU}
U(1/z)=U(z),  
\end{equation}
 and can be  analytically continued from the circle $S_1$  into its neighborhood $ \Gamma$. 

$U(z)$ can be expressed directly in terms of function $\phi(z)$:
\begin{equation} \label{Uphi}
U(z)= \int_{S_1}\frac{dz'}{2\pi i}\,\phi(z')\, \frac{\epsilon(z)-\epsilon(z')}{z'-z}.
\end{equation}
Note, that the integrand in (\ref{Uphi}) is regular at $z'\to z$.

Let us prove this useful relation,  following ideas of  Vekua \cite{Vek45}.
We start from the Cauchy formula
\begin{equation} \label{Ca}
U(z)=\int_{\partial\Gamma} \frac{d t}{2 \pi i}\, \frac{U(t)}{t-z},
\end{equation}
where $|z|=1$, and $\partial \Gamma=S_{1+\delta}-S_{1-\delta}$, $S_{1\pm \delta}$ denote circles of radius $1\pm \delta$
passed counter clockwise. Substitution of (\ref{U}) into (\ref{Ca}) yields
\begin{eqnarray} \label{Ca1}
U(z)=\int_{S_{1+\delta}} \frac{d t}{2 \pi i}\, \frac{[\epsilon(t)-\lambda]\,g_-(t)-\mu\, t \,g_-'(t)}{t-z}-\\ \nonumber
\int_{S_{1-\delta}} \frac{d t}{2 \pi i}\, \frac{[\epsilon(t)-\lambda]\,g_+(t)+\mu\, t \,g_+'(t)}{t-z}.
\end{eqnarray}
Several terms   vanish in (\ref{Ca1}) after integration, since $g_-(t)$ is analytical at $|t|>1$, 
and $g_+(t)$ is analytical at $|t|<1$. After substitution of (\ref{gpm}) in (\ref{Ca1}), the result reads as 
\begin{equation*} 
U(z)=\int_{\partial\Gamma} \frac{d t}{2 \pi i}\, \frac{\epsilon(t)}{t-z}\int_{S_1} 
\frac{d z'}{2 \pi i}\,\frac{\phi(z')}{z'-t}.
\end{equation*}
Change of the order of integration  leads to (\ref{Uphi}), 
since
\begin{equation*}
\int_{\partial \Gamma} \frac{d t}{2 \pi i}\, \frac{\epsilon(t)}{(t-z)(z'-t)}=\frac{\epsilon(z)-\epsilon(z')}{z'-z}
\end{equation*}
for $z,z'\in S_1$.
 
It is helpful to inverse relation (\ref{U}) and to express $g_+(z)$ in terms of $U(z)$: 
\begin{equation} \label{gpl}
g_+(z)=\int_{z_0}^z \frac{dz'}{\mu z'}\,U(z')\exp\Bigg\{\frac{i}{\mu}\,[{\mathcal F}(z')-{\mathcal F}(z)]\Bigg\},
\end{equation}
where
\begin{equation}
i\, {\mathcal F}(z)=\int_1^z \frac{dt}{t}\,[\epsilon(t)-\lambda]. \label{Fdef}
\end{equation}
The choice  of the integration path and its initial point $z_0$ in (\ref{gpl}) depends on $\epsilon(z)$, as
 will be discussed later. 
The general relations described here will be used in subsequent sections in analysis of equation (\ref{ieq})
for particular choices of function $\epsilon(z)$. In Section \ref{Toy} we shall describe two 
exact solutions of (\ref{ieq})
for rational $\epsilon(z)$. In Section \ref{Sec7} we perform a perturbative analysis in the limit $\mu\to0$
of equation (\ref{ieq}) with  $\epsilon(z)$ given by (\ref{eis}).  
\section{Toy models \label{Toy} }      
Equation (\ref{ieq}) can be solved exactly, if $\epsilon(z)$ is rational. In this case $U(z)$
is  also  rational, as follows immediately  from (\ref{U}). Furthermore, poles of $U(z)$ in the complex
$z$-plane can be located only at the poles of $\epsilon(z)$. This information is sufficient to obtain the 
explicit solution of equation (\ref{ieq}).
In the present  Section we describe this procedure in more detail for two rational  $\epsilon(z)$:
\begin{enumerate} 
\item  $\epsilon(z)=\epsilon_1(z)= -\frac{1}{2}(z+z^{-1})$, \label{ep1}
\item $  \epsilon(z)=\epsilon_2(z)= -(z+z^{-1})-\rho\,(z^2+z^{-2})$. \label{ep2}
\end{enumerate}
The reason for studying such  exactly solvable ``toy models'' is twofold. First, on the exact solutions one can 
check the accuracy of perturbative methods. Second, function  (\ref{eis}),  which arises in the Ising spin chain problem,
can be  approximated rather well  by the function $[B_0+B_1\,\epsilon_2(z)]$ with certain constants $B_0,\,B_1$, 
if $h_x$ is not close to the quantum phase transition point. 
\subsection{Toy model 1}
In this subsection we consider a toy model, in which free fermions at $h_z=0$ have the  dispersion law 
$\omega(\theta)=- \cos \theta$. Then function (\ref{eps}) reduces to 
\begin{equation} \label{ept1}
\epsilon(\theta;\Theta)=-2\, \cos( \Theta/2)\,\cos \theta, 
\end{equation}
and  equation (\ref{ieqA}) takes the form:
\begin{equation} \label{iet1}
[(-z-z^{-1})\cos(\Theta/2)-E(\Theta)]\,\phi(z)=-  \chi \,z \dashint_{S_1}\frac{dz'}{\pi i}\, \frac{\phi(z')}{(z'-z)^2}.
\end{equation}
After rescaling
\begin{equation}
{ \lambda}=\frac{E(\Theta)}{ 2 \cos(\Theta/2)}, \quad { \mu}=\frac{\chi}{2 \cos(\Theta/2) }, \label{resc}
\end{equation}
 we come to equation
(\ref{ieq}) with $\epsilon(z)=\epsilon_1(z)$. 
\begin{figure}[ht]
\centering
\includegraphics[width=.7 \linewidth]{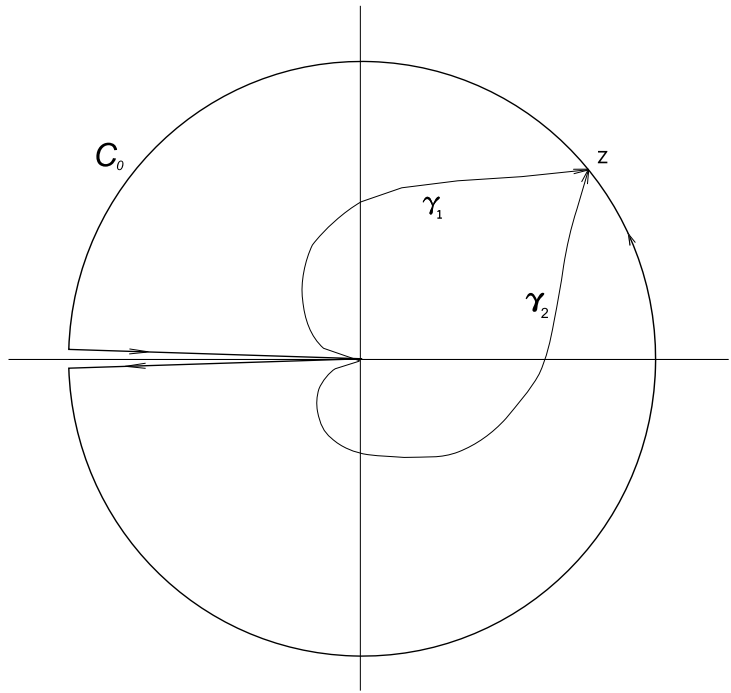} 
\caption{\label{Fig4} Integration paths in the $z'$-plane for the first toy model: 
arcs $\gamma_1$ and $\gamma_2$ going from $z_0=0$ to $z$ relate 
to (\ref{gpl});  loop $C_0$ is the integration path in  (\ref{Con1}).  
}
\end{figure}
\begin{figure}[ht]
\centering
\includegraphics[width=.7 \linewidth]{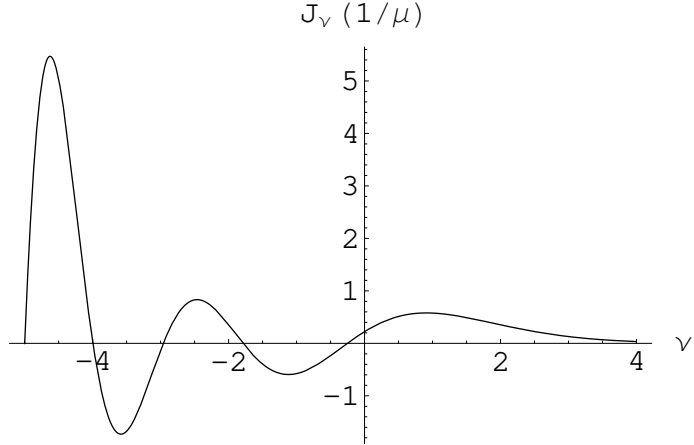} 
\caption{\label{Fig5} Plot of Bessel function $J_{\nu}(1/\mu)$ versus $\nu$ at $\mu=0.5$. 
}
\end{figure}
It is clear from (\ref{U}), that $U(z)$ is analytical in the whole complex plane $z$ for such $\epsilon(z)$. 
Indeed, the left-hand side of  (\ref{U}) is  analytical inside the circle $|z|=1$, since 
$g_+(z)$ is analytical there, and $\epsilon(z)$ is also analytical at $|z|<1$ except from the simple pole 
in the origin $z=0$, where  $g_+(z)$ takes zero value. Similarly, the right-hand side of  (\ref{U})
is analytical outside the circle $|z|=1$. Therefore $U(z)$ is a constant, and we can put $U(z)\equiv1$.
The function $g_+(z)$ is  determined by  (\ref{gpl}), with 
\begin{equation} \label{UF}
z_0=0, \quad U(z)\equiv1, \quad i\, {\mathcal F}(z)= -\lambda \log z-\frac{1}{2} (z-z^{-1}).
\end{equation}
We have set the  initial integration point  $z_0$ to zero  in (\ref{gpl}) to provide $g_+(0)=0$. 
Further, integration path in $z'$-plane should approach the point $z_0 =0$ from the left half-plane
${\rm Im}\, z'<0$ to guarantee convergence of the integral in $z'$ in the origin. Two integration 
paths $\gamma_1$ and $\gamma_2$, which satisfy all above conditions, are shown in Fig. \ref{Fig4}. Integration 
in (\ref{gpl}) along
each  path, either  $\gamma_1$ or $\gamma_2$,  should define {\it the same function} $g_+(z)$. 
This requirement  will be satisfied, if
\begin{equation} I(\lambda)=0, \label{Con1}
\end{equation}
 where 
\begin{equation} 
I(\lambda)=\int_{C_0} \frac{dz'}{ z'}\exp\Bigg\{\frac{i}{\mu}\,{\mathcal F}(z')\Bigg\}, \label{ila}
\end{equation}
and  the integration path $C_0= \gamma_2-\gamma_1$  is shown in Fig. \ref{Fig4}. Condition (\ref{Con1}) determines the
spectrum $\lambda$ for our first Toy model. 
After change of the integration variable $z'=u^{-1}$, the right-hand side in (\ref{ila}) reduces (up to the sign) 
to the integral, which gives the  well-known representation of  the Bessel function, 
\begin{equation}   
 \int_{-\infty}^{(0^+)}\frac{du}{u^{\nu +1}}\exp [a(u- u^{-1})/2]=2 \pi i\, J_\nu (a ) ,\label{Bess}
\end{equation}
with $\nu=-\lambda/\mu$, $a=1/\mu$. Integration contour in (\ref{Bess}) ``starts at infinity on the 
negative real $u$-axis, encircles the origin counter-clockwise, and returns to its starting point'', 
see formula  (5) in page 15 in \cite{Bat}.
So, the eigenvalues $\lambda_n$ of  equation (\ref{ieq}) with $\epsilon(z)=\epsilon_1(z)$ are determined
by zeros $\nu_n$ of the Bessel function $J_\nu (1/\mu )$  as the function of its order $\nu$:
\begin{equation} \label{ZBes} 
\lambda_n=-\nu_n\,\mu,\quad
J_{\nu_n}(1/\mu)=0. 
\end{equation}
A plot of  $J_{\nu}(1/\mu)$ versus $\nu$ at $\mu=0.5$ is shown in Fig. \ref{Fig5}. 
For the  energy spectrum $E_n (\Theta)$ in the first Toy model equations (\ref{ZBes}),  (\ref{resc})  yield 
\begin{eqnarray} \label{Ebes} 
&&E_n(\Theta)=-\nu_n\, \chi, \\
&& {\rm where}\;\;J_{\nu_n}\Bigg[\frac{2\cos(\Theta/2)}{\chi}\bigg]=0,\;\; n=1,2,\ldots.  \nonumber
\end{eqnarray}
Fig. \ref{Fig6}  shows the lowest  dispersion curves $E_n (\Theta)$ for $\chi=0.5$ and $\chi=0.2$. 
Note, that $E_n (\pi)= n \,\chi$ in this model. 

\begin{figure}
\includegraphics[width=.49\textwidth]{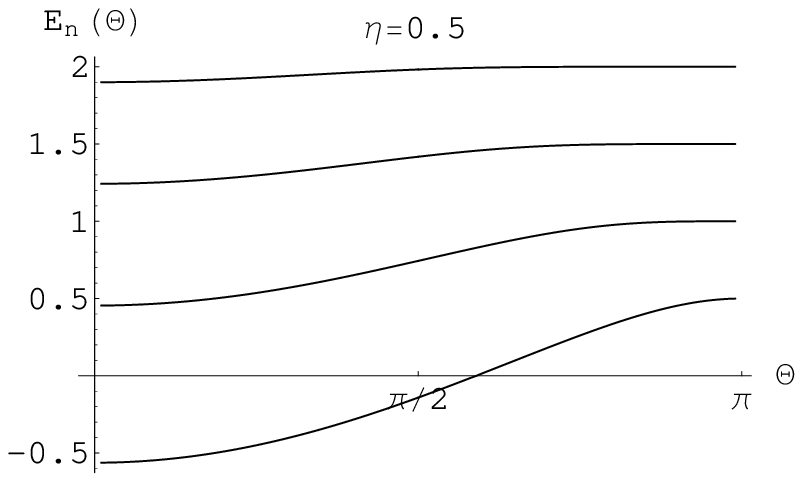}%
\hspace{.1in}%
\includegraphics[width=.49\textwidth]{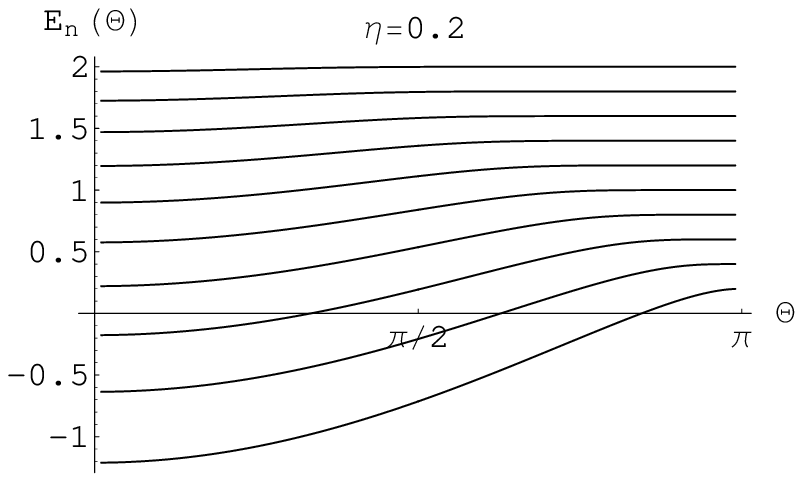}
\caption{Dispersion curves $E_n (\Theta)$ for the first toy model for $\chi=0.5$ (left) and $\chi=0.2$ 
(right) according to (\ref{Ebes}). }\label{Fig6}
\end{figure}

It is straightforward to extract the small-$\mu$ asymptotics from above exact results. In the $\mu\to+0$ limit the 
integral in (\ref{Con1}) is determined by the saddle points $z_{\pm}=-\lambda\pm i\sqrt{1-\lambda^2}$ of the function 
${\mathcal F}(z')$ . They lie on the circle $|z'|=1$, if  $|\lambda|<1$. 

If $|\lambda|$ is far enough from 1, they provide two separate 
complex conjugate contributions to the integral in (\ref{ila}). In the leading order in $\mu$,  this yields
\begin{equation} \label{qrule1a}
\lambda_n \theta_a+ \sin\theta_a=\pi\mu\bigg(n-\frac{1}{4}\bigg),\quad \cos \theta_a=-\lambda_n,
\end{equation}
in agreement with semiclassical quantization formula (\ref{qrule}). 

If $\lambda$ approaches  $-1$, two saddle points $z_{\pm}$ merge at $1$, and the integral (\ref{ila}) 
becomes proportional to the Airy function \cite{AbrSt}
\begin{equation}
 I(\lambda) \approx \int_{-\infty}^\infty d\theta \exp\bigg\{\frac{i}{\mu}
\bigg[-(\lambda+1)\,\theta+\frac{\theta^3}{6}\bigg]\bigg\}=
2 \pi (2 \mu)^{1/3} {\rm Ai}[-(\lambda+1)\,(2 )^{1/3}\,\mu^{-2/3}]
\end{equation}
in the limit $\mu\to 0$.
From (\ref{Con1}) we get  in this case
\begin{equation*}
\lambda_n=-1+ \mu^{2/3} \,2^{-1/3}\,z_n,\;\; {\rm where} \quad {\rm Ai}[-z_n]=0,\quad n=1,2,\ldots,
\end{equation*}
in agreement with (\ref{ei}).

If $\lambda$ gets close to $+1$,  $z_{\pm}$ approach  $-1$. 
However, there are still two  saddle point contributions to $I(E)$, coming from the 
the upper and lower edges of the contour $C_0$ at $z'=-1$, see Fig. \ref{Fig4}.
The resulting equation for $\lambda$ reads
\begin{equation} \label{AB}
\tan\bigg(\frac{\pi \lambda}{\mu}\bigg)=-\frac{ {\rm Ai}[(\lambda-1)\,(2/\mu^2 )^{1/3}]}
{ {\rm Bi}[(\lambda-1)\,(2/\mu^2 )^{1/3}]},
\end{equation}
where ${\rm Bi}(x)$ is the second Airy functions, which  
solves (together with  ${\rm Ai}(x)$) the differential equation equation $y''(x)-x \,y(x)=0$,  and has the 
following integral representation \cite{AbrSt}
\begin{equation}
(3a)^{-1/3}\pi \,{\rm Bi}[\pm (3a)^{-1/3}\,x]=\int_0^\infty dt\, [\exp(-a t^3\pm x t)+\sin(a t^3\pm x t)].
\end{equation}

Equation (\ref{AB}) describes the  crossover  at $\lambda\approx 1$ of the $\mu\to0$ asymptotics from the 
semiclassical regime (\ref{qrule1a})  at $\lambda<1$,  to   $\lambda_n\approx \mu\,n$  at $\lambda>1$.  
\subsection{Toy model 2}
In the first Toy model with $\omega(\theta)=-\cos\theta$,  variation of the 
total momentum $\Theta$ only rescales $\epsilon(\theta;\Theta)$, see (\ref{ept1}).
However, for the real free-fermion dispersion law (\ref{fdis}), the topology of 
$\epsilon(\theta;\Theta)$ also changes with increasing  $\Theta$, compare Figs. \ref{F2} 
and \ref{F3}. To mimic this  property, let us  modify the ``free-fermion dispersion law'' to 
\begin{equation} \label{tru2}
\omega_2(\theta)=-\cos \theta -\gamma \cos(2\, \theta). 
\end{equation}
with positive $\gamma$.
Then the function (\ref{eps}) takes the form 
\begin{equation} \label{epp2}
\epsilon(\theta;\Theta)=-2\, \cos( \Theta/2)\,\cos \theta-2\,\gamma\,\cos( \Theta)\,\cos(2\, \theta). 
\end{equation}
Equation (\ref{ieqA}) reads now 
\begin{eqnarray} \label{iet2}
&&\{[-(z+z^{-1})\cos(\Theta/2)- (z^2+z^{-2})\,\gamma\,\cos \Theta]-
E(\Theta)]\}\,\phi(z)\\
&&=-  \chi \,z \dashint_{S_1}\frac{dz'}{\pi i}\, \frac{\phi(z')}{(z'-z)^2}. \nonumber
\end{eqnarray}
After rescaling
\begin{equation}
{ \lambda}=\frac{E(\Theta)}{  \cos(\Theta/2)}, \quad { \mu}=\frac{\chi}{ \cos(\Theta/2) },
\quad { \rho}=\frac{\gamma\,\cos\Theta}{ \cos(\Theta/2) }, \label{resc3}
\end{equation} 
it reduces to (\ref{ieq}) with $\epsilon(z)=\epsilon_2(z)= -(z+z^{-1})-\rho\,(z^2+z^{-2})$.

One can naturally come to this Toy model  approximating   the exact free-fermion dispersion law (\ref{fdis})
by three initial terms of its Fourier expansion. This approximation is rather good, if $h_x$ is not too close
to the quantum phase transition point $h_x=1$.

It follows from (\ref{U}), that $U(z)$  can be now written as
\begin{equation*} 
U(z)=A_1 +A_2\, (z+z^{-1}),
\end{equation*}
with two constants $A_1,\,A_2$. 
The function $g_+(z)$  determined by (\ref{gpl})  takes the form
\begin{equation} \label{gpl2}
g_+(z)=\int_{0}^z \frac{dz'}{\mu\, z'}\,\bigg[A_1 +A_2\, \bigg(z'+\frac{1}{z'}\bigg)\bigg]
\exp\Bigg\{\frac{i}{\mu}\,[{\mathcal F}(z')-{\mathcal F}(z)]\Bigg\},
\end{equation}
where
\begin{figure}[t]
\centering
\includegraphics[width=1 \linewidth]{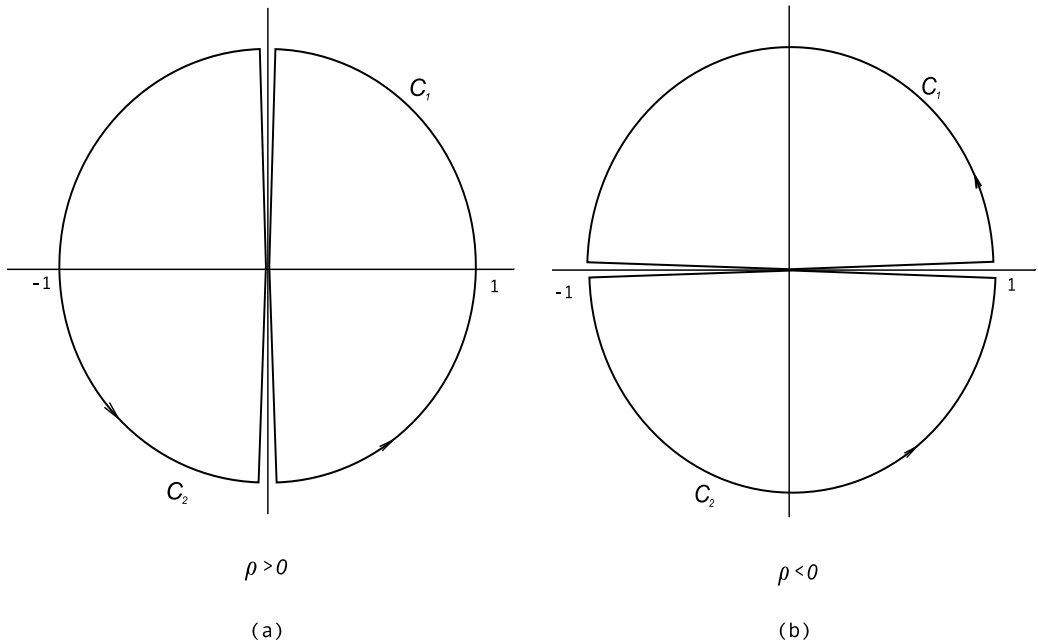} 
\caption{Integration contours $C_1$ and $C_2$ in (\ref{W}): (a) for $\rho>0$,
(b)  $\rho<0$ . Both $C_1$ and $C_2$ start and finish in the origin.}\label{Fig7}
\end{figure}
\begin{equation*} 
i {\mathcal F}(z)=-\lambda \log z -\bigg(z-\frac{1}{z}\bigg)-\frac{\rho}{2} \,\bigg(z^2-\frac{1}{ z^2}\bigg)
\end{equation*}
To provide convergence of the integral in (\ref{gpl2}), the variable $z'$ should approach the origin  inside 
one of two allowed sectors, which are  defined by the condition ${\rm Re}(\rho/z'^2)<0$, or explicitly
\begin{eqnarray*}
\pi/4&<&\arg z'<3 \pi/4, \quad{\rm and } \;\;\; -3\pi/4<\arg z'<-\pi/4,\quad\, {\rm if}\;\; \rho >0,\\
-\pi/4&<&\arg z'<\pi/4, \quad\;\,{\rm and } \quad\quad 5\pi/4<\arg z'<3\pi/4,\quad\;\; {\rm if}\;\; \rho <0.
\end{eqnarray*}
Definition (\ref{gpl2}) becomes  unambiguous and independent on the integration path between the
points $0$ and $z$, if the constants $A_1$ and $A_2$ solve two linear uniform equations:
\begin{eqnarray*}
 W_{11}(\lambda)\,A_1+W_{12}(\lambda)\,A_2=0, \\
 W_{21}(\lambda)\,A_1+W_{22}(\lambda)\,A_2=0 ,
\end{eqnarray*}
where 
\begin{eqnarray} \label{W}
W_{ij}(\lambda)&=&\int_{C_i} \frac{dz'}{ z'}\,w_j(z') 
\exp\bigg[\frac{i}{\mu}\,{\mathcal F}(z')\bigg] , \quad i,j=1,2, \\
w_1(z')&=&1, \quad w_2(z')=z'+\frac{1}{z'}. \nonumber
\end{eqnarray} 
Integration contours $C_{1}$ and $C_{2}$ for the two cases 
 $\rho>0$ and $\rho<0$  are shown in Fig. \ref{Fig7}.  
We fix the  branch of ${\mathcal F}(z')$ in (\ref{W}) by the condition ${\mathcal F}(1)=0$.

Eigenvalues $\lambda_n$ in the second Toy model  are then determined  by the requirement:
\begin{equation*}
W_{11}(\lambda_n) W_{22}(\lambda_n)-W_{12}(\lambda_n) W_{21}(\lambda_n)=0.
\end{equation*} 
With known $\lambda_n$, we  obtain finally $E_n(\Theta)$ by use of (\ref{resc3}).  
\\*{\bf Note.} In the same way one can find exact solutions of equation (\ref{ieq}) 
with $\epsilon(z)=\sum_{k=1}^K c_k (z^k+z^{-k})$. In this case the  eigenvalues  $\lambda_n$ of (\ref{ieq}) 
are determined from $\det  W(\lambda_n)=0$, where  $ W(\lambda)$ denotes a $K\times K$-matrix.

The small-$\mu$ behavior of the second Toy model is more rich than in the previous case. It is 
determined by the  location of four saddle points of the integrals in (\ref{W}). These saddle points  are 
the solutions of the equation $\epsilon_2(z)=\lambda$.
We skip further discussions of the $\mu\to 0$ asymptotics of the Toy model 2, since it is very similar to 
that of original equation (\ref{ieqA}), which will be  described in the following Section.
\section{Weak coupling expansion}  \label{Sec7}
In this section we return to the original equation (\ref{ieqA}), describing bound-spinons 
in the two-fermion approximation, and obtain its perturbative solution in the weak coupling
limit $\chi\to 0$. Several perturbative schemes have been developed for the analogous problem 
in  IFT \cite{FonZam2003,FZ06,Rut05}. Here we shall use a different procedure, which
utilizes  results of Section \ref{Sec5}.
\begin{figure}[ht]
\centering
\includegraphics[width=.6 \linewidth]{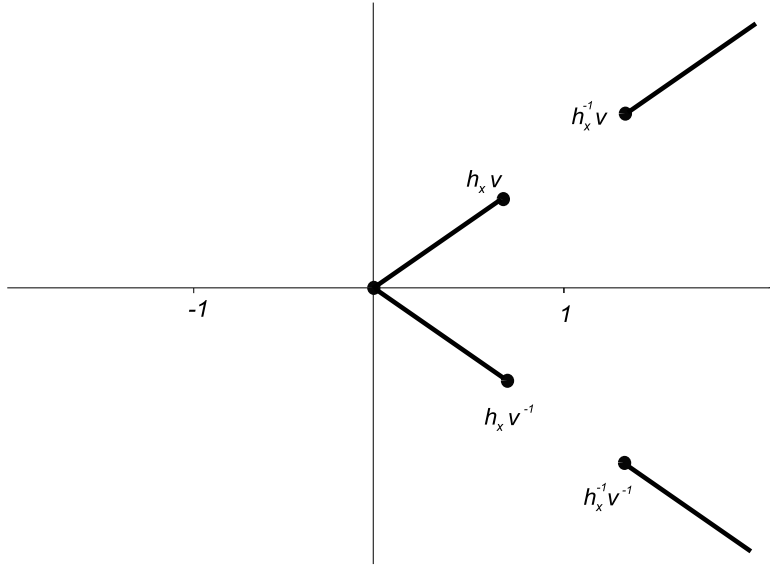} 
\caption{Square root branching points of function $\epsilon(z)$ determined  by (\ref{not}),
and cuts on $z$-plane.}\label{Fig8}
\end{figure}

First, let us rewrite (\ref{ieqA})  in notations of Section \ref{Sec5}:
\begin{equation} 
[\epsilon(z)-\lambda]\,\phi(z)=-  \mu\,z \dashint_{S_1}\frac{dz'}{\pi i}\, \frac{\phi(z')}{(z'-z)^2},\label{ie3}
\end{equation}
where 
\begin{eqnarray} \label{not}
\epsilon(z)&=&2\sqrt{h_x}\Bigg\{\Bigg[h_x+\frac{1}{h_x}- zv -\frac{1}{zv}\Bigg]^{1/2}+
\Bigg[h_x+\frac{1}{h_x}- \frac{z}{v} -\frac{v}{z}\Bigg]^{1/2}\Bigg\}, \\
\mu&=&\chi, \quad \lambda= E(\Theta),\quad v=\exp(i\Theta/2). \label{no1}
\end{eqnarray}
Remind, that $S_1$ is the unit circle, and $\phi(1/z)=-\phi(z)$. 

Function $\epsilon(z)$ has six branching points $0, h_x v, h_x v^{-1},h_x^{-1} v,h_x^{-1} v^{-1}, \infty $.
Its Riemann surface $\mathcal L$ has four sheets ${\mathcal L}_{\alpha, \beta}$ with $\alpha, \beta=+,-$,
which are distinguished by the signs of the first an the second terms in braces in (\ref{not}) at $z=1$.
To separate the sheets, we draw four cuts in the $z$-plane shown in Fig. \ref{Fig8}. Equation (\ref{ie3}) is written in
the sheet ${\mathcal L}_{+, +}$, which we shall call ``the physical sheet''. 

Equations (\ref{gpm}), (\ref{U}) define functions $g_{\pm}(z)$ and $U(z)$ in ${\mathcal L}_{+, +}$. It follows
from these equations and  (\ref{not}), that all singularities of $U(z)$ in the physical sheet  are the square root
singularities of $\epsilon(z)$. Therefore, $U(z)$ is finite in ${\mathcal L}_{+, +}$. 

Let us turn now to the integral formula (\ref{gpl}). As in the previous Section, we should put $z_0=0$ in it to 
provide $g_+(0)=0$. The function ${\mathcal F}(z)$, determined by (\ref{Fdef}), 
is singular in the physical sheet at $z\to 0$, 
${\mathcal F}(z)\sim z^{-1/2}$. Nevertheless, the integral in (\ref{gpl}) converges, if the 
whole integration path lies in ${\mathcal L}_{+, +}$, approaching to the origin either from the right, or from the 
left side. This follows from the fact, that $\epsilon(z)$ is positive at real $z$  in the physical sheet.

Furthermore, three integral equalities must be satisfied
\begin{equation} 
\int_{C_\alpha} \frac{dz}{ z}\,U(z)\,\exp\bigg\{\frac{i}{\mu}\,{\mathcal F}(z)\bigg\}=0, \quad \alpha=0,1,2, \label{cond}
\end{equation}
where the integration path $C_{0}$ is shown in Fig. \ref{Fig4}, and  paths $C_{1,2}$ are drawn in Fig. \ref{Fig7}(b).
Note, that  only two  conditions in (\ref{cond}) are independent, since $C_0=C_1+C_2$.
Conditions (\ref{cond}) are necessary to guarantee, that:
\begin{itemize}
\item the integral in (\ref{gpl}) defines a single valued function in the unit circle $S_1$,
\item  $g_+(z)$ given  by (\ref{gpl}) vanishes, when $z$ approaches to the origin in ${\mathcal L}_{+, +}$
either from the left, or from the right side of the cuts, see Fig. \ref{Fig8}.
\end{itemize}
In the limit  $\mu\to+0$  the integrals in (\ref{cond}) are determined by contributions of the saddle point 
of function ${\mathcal F}(z)$, i.e. by the solutions of equation 
\begin{equation} \label{spe}
\epsilon(z)=\lambda.
\end{equation} 
This equation has four solutions $z_a,z_b, z_a^{-1},z_b^{-1}$ in the Riemann surface $\mathcal L$,
which  are given  by
\begin{eqnarray*} 
z_\alpha+z_\alpha^{-1}&=& \frac{\lambda^2\cos(\Theta/2)\pm\{[\lambda^2-16\sin^2(\Theta/2)]
[\lambda^2-16\,h_x^2\,\sin^2(\Theta/2)]\}^{1/2}}{8\,h_x\,\sin^2(\Theta/2)}, \\
 \alpha&=&a,b.
\end{eqnarray*}
 Important for us   are the saddle points, which lie on the circle $|z|=1$, or are close to it.  
In the simplest case, there are only two such solutions of (\ref{spe}):
$z_a=\exp(i\theta_a)$, and $z_a^{-1}=\exp(-i\theta_a)$. Suppose, they  are well separated, being 
 far enough from 1 and -1. Then the  saddle point 
expansion of the integral in (\ref{cond}) over the path $C_0$ leads to the asymptotic expansion
\begin{equation}
-{\mathcal F}(\theta_a)=\mu \pi \bigg(n-\frac{1}{4}\bigg)+\mu\, \arg \bigg\langle
U(\theta_a+ \delta \theta) \exp\bigg[\frac{i}{\mu}
\Delta\mathcal{F}(\theta_a+\delta \theta)\bigg]\bigg\rangle,  \label{reqs}
\end{equation}
valid to all orders in $\mu$. 
Here we use short notations ${\mathcal F}(\theta)={\mathcal F}[z(\theta)]$, 
$U(\theta)= U[z(\theta)]$, and  
$$
\Delta\mathcal{F}(\theta_a+\delta \theta)=\mathcal{F}(\theta_a+\delta \theta)-\mathcal{F}(\theta_a)-
 \frac{\mathcal{F}''(\theta_a)}{2}\, \delta \theta^2.
$$
Averaging  $ \langle f(\delta \theta) \rangle $ implies formal term-by term integration
$$
 \langle f(\delta \theta)\rangle= \frac{\int_{-\infty}^\infty d\,\delta \theta \, f(\delta \theta) 
 \exp\bigg[\frac{i\mathcal{F}''(\theta_a) }{2\,\mu}\,\delta \theta^2\bigg] }{
 \int_{-\infty}^\infty d\,\delta \theta \, 
\exp\bigg[\frac{i\mathcal{F}''(\theta_a) }{2\,\mu}\,\delta \theta^2 \bigg]},
$$
with   $f(\delta \theta)$ expanded into a power series in $\delta \theta$. Under appropriate normalization
of $\phi(z)$, the function $U(\theta_a+ \delta \theta)$ can be expanded as 
\begin{equation} \label{Upow}
U(\theta_a+ \delta \theta)=1+\sum_{i=1}^{\infty}\sum_{l=0}^\infty c_{il}\,\mu^i\,\delta \theta^l.
\end{equation}
To obtain the coefficients $c_{il}$ explicitly, one should write down the  Neumann series, 
giving the formal  solution of  
equation (\ref{ie3}) in the class of generalized functions, 
\begin{eqnarray}
\phi(\theta)&=& \phi^{(0)}(\theta) +\phi^{(1)}(\theta)+O(\mu^2), \label{exphi}\\
 \phi^{(0)}(\theta)&=&C \,[\delta(\theta-\theta_a)-\delta(\theta+\theta_a)], \nonumber \\
 \phi^{(1)}(\theta)&=& \frac{\mu \,C}{4\,\pi\,[\epsilon(\theta)-\lambda]}
\bigg\{\frac{1}{\sin^2[(\theta-\theta_a)/2]}-\frac{1}{\sin^2[(\theta+\theta_a)/2]}\bigg\}. \label{psi1}
\end{eqnarray}
Substitution of this expansion into (\ref{Uphi}) gives us a smooth function $U(\theta)$
as a power series in $\mu$
\begin{eqnarray}
U(\theta)&=& U^{(0)}(\theta) +U^{(1)}(\theta)+O(\mu^2), \label{exU}\\
 U^{(0)}(\theta)&=&\frac{C}{2 \pi}\, [\epsilon(\theta)-\epsilon(\theta_a)]\,
\bigg[\frac{1}{1-\exp[i(\theta- \theta_a)]}-\frac{1}{1-\exp[i(\theta+ \theta_a)]}    \bigg], \nonumber
\end{eqnarray}
which in turn can be expanded in $\theta$  at $\theta=\theta_a$. 
The result is brought  to the form (\ref{Upow}), if one  puts  $C=-2\pi i/\epsilon'(\theta_a)$.

The  perturbation procedure described above  should be modified, if equation (\ref{spe}) has four solutions
$z_a,z_b,z_a^{-1},z_b^{-1}$ on  the circle $|z|=1$, $z_{a,b}=\exp[i\theta_{a,b}]$, see Fig. \ref{F3}. 
If they are well separated, 
 the zero-order function $\phi^{(0)}(\theta)$
 in (\ref{exphi}) can  be written as 
\begin{equation} \label{phi0}
\phi^{(0)}(\theta)=C_a \,[\delta(\theta-\theta_a)-\delta(\theta+\theta_a)]+
C_b \,[\delta(\theta-\theta_b)-\delta(\theta+\theta_b)].
\end{equation}
Using the leading saddle point approximation for the  integrals in (\ref{cond}) with $j=1,2$, we obtain
a system of two equations
\begin{eqnarray*} 
u_a \,\bigg[\frac{2 \pi \mu}{\epsilon'(\theta_a)}\bigg]^{1/2}e^{i \pi/4+i{\mathcal  F}(\theta_a)/\mu}+
u_b \,\bigg[\frac{2 \pi \mu}{-\epsilon'(\theta_b)}\bigg]^{1/2}\, e^{-i \pi/4+i{\mathcal F}(\theta_b)/\mu}=0, \\
u_a \,\bigg[\frac{2 \pi \mu}{\epsilon'(\theta_a)}\bigg]^{1/2}e^{-i \pi/4-i{\mathcal  F}(\theta_a)/\mu}+
u_b \,\bigg[\frac{2 \pi \mu}{-\epsilon'(\theta_b)}\bigg]^{1/2}\, e^{i \pi/4-i{\mathcal F}(\theta_b)/\mu}=0,
\end{eqnarray*}
for  constants $u_a=U^{(0)}(\theta_{a})$, $u_b=U^{(0)}(\theta_{b})$, which  are simply related with $C_a,\, C_b$.
Setting its determinant to zero, we get
\begin{equation} \label{gen2}
{\mathcal F}(\theta_b)-{\mathcal F}(\theta_a)=\pi \mu \bigg(n-\frac{1}{2}\bigg)+O(\mu^2),
\end{equation} 
in agreement with the semiclassical formula (\ref{qrule1}).

It is easy to determine the eigenvalues $\lambda_n$ of (\ref{ie3}) in the limit $\mu\to+0$ for 
$\lambda>\epsilon(\theta=\pi)$. Let us rewrite condition (\ref{cond}) with $\alpha=0$ as
\begin{eqnarray}\nonumber
&&0=\int_{-\pi}^\pi d \theta\, U(\theta)\,\exp[i{\mathcal F}(\theta)/\mu]+
\int_{\pi}^{\pi+i\infty} d \theta \,U(\theta)\,\exp[i{\mathcal F}(\theta)/\mu]+\\
\nonumber&&
\int_{-\pi-i\infty}^{-\pi} d \theta \,U(\theta)\,\exp[i{\mathcal F}(\theta)/\mu]=
\int_{-\pi}^\pi d \theta\, U(\theta)\,\exp[i{\mathcal F}(\theta)/\mu]+\\&&
\{1-\exp[-i{\mathcal F}(2 \pi)/\mu]\}\int_{\pi}^{\pi+i\infty} d \theta \,U(\theta)\,\exp[i{\mathcal F}(\theta)/\mu].
\label{lam2}
\end{eqnarray}
In the second equality we have taken into account, that $U(\theta-2\pi)=U(\theta)$, and
${\mathcal F}(\theta-2\pi)={\mathcal F}(\theta)-{\mathcal F}(2 \pi)$.

If $\lambda>\epsilon(\theta=\pi)$, the single saddle point $z_a$ in the integration path $C_0$ is located 
in the interval $(-1,0)$ on the real $z$-axis. Two contributions of this saddle point will cancel
each other in (\ref{lam2}), if ${\mathcal F}(2 \pi)=-2\pi n\,\mu$ with integer $n$, i.e. 
\begin{equation} \label{lam>}
\lambda_n=\frac{1}{2\pi}\int_0^{2 \pi}d\theta\,\epsilon(\theta)+\mu\, n.
\end{equation}
This asymptotic formula is valid to all orders{\footnote  {Corrections to (\ref{lam>})  of order
$\mu\,\exp[-A(\lambda)/\mu]$ can be obtained from (\ref{AB1}).}} in $\mu$.
Rewriting it in  original variables  (\ref{no1}) and using (\ref{eps}),  
one obtains the bound-spinon spectrum 
\begin{eqnarray} \label{E2gen}
&&E_n(\Theta)=\frac{1}{2\pi}\int_0^{2 \pi}d\theta\,2\omega(\theta)+ \chi\, n \\
&&\textrm{for $E_n(\Theta)>\epsilon(\pi,\Theta)$ and  $\chi\to+0$}. \label{colam}
\end{eqnarray}
This result has a clear physical interpretation. Condition (\ref{colam}) means, that  a bound-spinon can be 
considered semiclassically as a large enough domain, bounded by two domain walls. These domain walls  move 
back and forth without collisions according to the classical equations of motion corresponding 
to the Hamiltonian (\ref{Hcl}).   
The two terms on the right-hand side of (\ref{E2gen}) give the  kinetic  and 
potential energy of domain walls averaged over  their oscillations. So, the energy of 
such a  bound-spinon does not depend on its momentum $\Theta$, and hence, it can not move as a whole
along the spin chain. 

The asymptotic formulas (\ref{reqs}), (\ref{gen2}), (\ref{lam>}) describe the small-$\mu$ behavior of eigenvalues
$\lambda_n$ of (\ref{ie3}) for generic values of $\lambda$ and $\Theta$. Five crossover regimes are realized,
when $\lambda$ approaches the  critical values of $\epsilon(\theta)$ and two or four solutions of equation
(\ref{spe}) merge.  All these regimes can be analyzed by means of almost the same perturbation procedure 
utilizing equations (\ref{Uphi}) and (\ref{cond}). Minor modifications in it are caused by further degeneracy
of  ${\mathcal F}(\theta)$ near  the saddle point $\theta_0$: ${\mathcal F}(\theta)-{\mathcal F}(\theta_0)\sim
(\theta-\theta_0)^j$ where $j=3$ or $j=5$ instead of $j=2$ in the  case 
of Morse saddle point considered previously. In the 
$\mu\to 0$ limit this leads to perturbation expansions for $\lambda_n$  
in  powers of parameter $\mu^{1/3}$ or $\mu^{1/5}$. Below we present the leading terms of 
these expansions, skipping the details of calculations. 
\begin{enumerate}
\item $\theta_a\approx 0$, $0<\Theta<\Theta_m$, where $\Theta_m=2  \arccos h_x$, 
\begin{equation} \label{A0}
\lambda_n=\epsilon(0)+z_n\, \mu^{2/3} \,[\epsilon''(0)/2]^{1/3}.
\end{equation}
\item  $\theta_a\approx\pi$, $0<\Theta<\pi$,
\begin{equation} \label{AB1}
\tan\bigg[\frac{{\mathcal F}(\pi)}{\mu}\bigg]=
\frac{ {\rm Ai}\Big[[\lambda_n-\epsilon(\pi)]\,[-2/\epsilon''(\pi) ]^{1/3}\mu^{-2/3}\Big]}
{ {\rm Bi}\Big[[\lambda_n-\epsilon(\pi)]\,[-2/\epsilon''(\pi) ]^{1/3}\mu^{-2/3}\Big] }.
\end{equation}
\item  $\Theta_m<\Theta<\pi$
\begin{enumerate}
\item $\theta_a\approx \theta_b\approx \theta_0>0$, where $\cos\theta_0=\frac{\cos(\Theta/2)}{\cos(\Theta_m/2)}$.
\begin{eqnarray}
\lambda_{n,1}=\epsilon(\theta_0)+z_n\, \mu^{2/3} \,[\epsilon''(\theta_0)/2]^{1/3},\\
\lambda_{n,2}=\epsilon(\theta_0)+z_n'\, \mu^{2/3} \,[\epsilon''(\theta_0)/2]^{1/3}, 
\end{eqnarray}
where  $n=1,2,\ldots$.
\item $\theta_b\approx0$, $\theta_a>0$,
\begin{equation} \label{AB2}
{\rm cotan}\bigg[-\frac{{\mathcal F}(\theta_a)}{\mu}-\frac{\pi}{4}\bigg]=
\frac{ {\rm Ai}\Big[[\lambda_n-\epsilon(0)]\,[-2/|\epsilon''(0) |]^{1/3}\mu^{-2/3}\Big]}
{ {\rm Bi}\Big[[\lambda_n-\epsilon(0)]\,[-2/|\epsilon''(0) |]^{1/3}\mu^{-2/3}\Big] }.
\end{equation}
\end{enumerate}
\item  $\Theta= \Theta_m$, $\theta_a\approx0$, $ \theta_b\approx i \theta_a$,
\begin{equation}
\lambda_n=\bigg\{\epsilon(\theta)+\mu^{4/5}\,\bigg[\frac{\partial^4 \epsilon(\theta)/\partial\theta^4}{6}\bigg]^{1/5}
\,c_n\bigg\}|_{\theta=0},
\end{equation}
where $\epsilon(0)=2\, (1-h_x^2)^{1/2}$, $\partial^4 \epsilon(\theta)/\partial\theta^4|_{\theta=0}=
h_x^2/\sqrt{1-h_x^2}$, and     $n=1,2,\ldots$.
\end{enumerate}
The numbers $z_n$ and $z_n'$  are consecutive zeroes of ${\rm Ai}(-z)$ and ${\rm Ai}'(-z)$ respectively, and  $c_n$ 
are the solutions of equation
\begin{eqnarray}
&&\int_0^\infty dy \,  \bigg[\sin\bigg( \frac{y^5}{20}-y\, c_n\bigg)-
\exp\bigg(- \frac{y^5}{20}+y \,c_n\bigg)\bigg]\cdot \\&&\nonumber
\int_0^\infty dx \, x^2 \cos\bigg( \frac{x^5}{20}-x \,c_n\bigg)=
\int_0^\infty dx \,  \cos\bigg( \frac{x^5}{20}-x \,c_n\bigg)\cdot \\
&&\int_0^\infty dy \, y^2 \bigg[\sin\bigg( \frac{y^5}{20}-y\, c_n\bigg)+
\exp\bigg(- \frac{y^5}{20}+y\ c_n\bigg)\bigg], \nonumber
\end{eqnarray}
$c_1=1.787$, $c_2=3.544$, $c_3=5.086$.

Note, that all asymptotic regimes described above are realized also in the second Toy model, and the four
regimes described by equations (\ref{reqs}), (\ref{lam>}), (\ref{A0}), (\ref{AB1}) are relevant to 
the first Toy model as well.
\section{Discussion} \label{Dis}
In this paper we apply a technique developed in IFT to study the effect the discreteness of  the  lattice 
 on the  kink confinement  in a non-critical one-dimensional system with explicitly 
broken  $\mathbb{Z}_2$-symmetry. We study the model of the Ising spin-$1/2$ chain ferromagnet at zero temperature 
in the presence of a skew magnetic field, as a particular realization of such a system. The magnetic 
field $h_x$,  transverse to the easy $z$-axis,  serves to induce the quantum phase transition at $h_x=\pm 1$ 
and to allow kinks in the ordered 
phase $|h_x|<1$ to move. The longitudinal field $h_z$ breaks the  $\mathbb{Z}_2$-symmetry and leads to confinement of 
kinks into pairs, the bound-spinons. We calculate the dispersion law $E_n(\Theta)$ of bound-spinons in the limit
of small $h_z$ in the leading order in this parameter.

The dispersion law of bound-spinons can be  understood within a simple heuristic picture, in which
one treats a bound-spinon as bound state of two classical particles (kinks)  
attracting each  other with a linear potential proportional to $h_z$. 
If these particles are close enough to each other, such that they can meat  during their classical motion, then 
they can drift  as a whole along the spin chain. 
In this case, one can determine the dispersion law $E_n(\Theta)$ of a  bound-spinon from the 
Bohr-Sommerfeld quantization rule, if $n\gg1$.  

On the other hand, large enough bound-spinons can be viewed as two independent well separated  kinks. 
The  linear  potential leads to localization of an isolated kink in the discrete spin chain, 
similarly to localization of an electron moving in a periodic 
potential by a uniform electric field \cite{Zim72}. The classical motion of isolated kinks looks like oscillations 
around certain positions in the chain. If two kinks forming a bound-spinon do not meet during their classical 
oscillations, the  velocity $v_n(\Theta)$ of the bound-spinon is  zero. Since 
$v_n(\Theta)=\partial E_n(\Theta)/\partial\Theta $, we conclude $E_n(\Theta)$ is constant for such a bound-spinon.
It is clear that the energy spectrum should be equidistant in this region 
\begin{equation} \label{diE}
E_{n+1}(\Theta)-E_{n}(\Theta)=2 h_z \bar{\sigma},
\end{equation}
since the bound-spinon state $|\Phi_{n+1}(\Theta)\rangle$ contains one extra inverted spin in the domain 
bounded by two kinks, compared to the state $|\Phi_n(\Theta)\rangle$. 

A more systematic theory is based on the perturbative analysis of the singular integral equation (\ref{ieqA}), 
which is analogous to the  Bethe-Salpeter equation in IFT . The equations have been obtained in the 
two-fermion approximation and become exact in the limit $h_z\to 0$. The bound-spinon energy spectrum $E_{n}(\Theta)$ 
is determined in this approximation as  eigenvalues of equation (\ref{ieqA}). 
We develop perturbation theory for this equation 
in small $h_z$ and, concentrating  on its leading order,  describe eight asymptotic regimes for the 
 bound-spinon energy spectra $E_{n}(\Theta)$, 
instead of two regimes  (\ref{low}) and (\ref{wkbe}) known in IFT. 

It was shown \cite{FZ06}, that  the IFT Bethe-Salpeter equation  reproduces 
 the energy spectra of stable ``mesons'' with reasonable accuracy not only  
in the limit $h\to 0$, but also at finite, and even at large values of the magnetic field $h$. 
If this situations holds in the discrete-lattice case, it would  be reasonable to study the 
spectrum of  equation  (\ref{ieqA}) at finite $h_z\sim 1$ as well. Few steps in this direction give 
exact solutions of its  simplified versions (\ref{iet1}) and (\ref{iet2}), which were described 
in Section \ref{Toy}.  

In conclusion let us comment on the higher order corrections to the obtained bound-spinon spectrum 
$E_{n}(\Theta)$ in the limit $h_z\to 0$.  It is  straightforward to determine the higher order terms in the 
weak coupling expansion of the eigenvalues of equation (\ref{ieqA}), as it was  described in Section \ref{Sec7}. 
However, one should also take into account corrections  to  equation (\ref{ieqA}), which is 
an approximation by itself.
First, the kernel $G_{\Theta}(\theta,\theta')$  in the integral in equation (\ref{bs2}) has a regular 
part $G_{\Theta}^{(reg)}(\theta,\theta')$  omitted in (\ref{ieqA}). Only few
modifications in the perturbation schema of Section \ref{Sec7} are needed  to apply it to equation (\ref{bs2})
with  this term restored. Second, starting from the second order in $h_z$ one should take into multi-fermion
corrections, which arise from the four-fermion, six-fermion, ...,  contributions to the
bound-spinon eigenvector $|\Phi_n(\Theta)\rangle$.  
As we know from IFT  \cite{FonZam2003,FZ06,Rut05}, 
multi-fermion effects lead to renormalization of the 
fermion dispersion law $\omega(\theta)$ and coupling  constant $2 h_z \bar{\sigma}$, and 
should lead to the decay of unstable bound-spinons.  All these effect are lost in the two-fermion 
approximation (\ref{eig2}), but contribute to the bound-spinon dispersion law  $E_{n}(\Theta)$
in  higher orders in $h_z$. 
 A perturbative analysis of multi-fermion effects
in the discrete  Ising model (\ref{Ham}) is an interesting problem, which remains
for future work.
\section{Acknowledgements} I am thankful to the Institute of Theoretical Physics
of the University of M\"unster for hospitality.
I would like to express my gratitude to Gernot M\"unster for many helpful discussions
and valuable remarks.  This work 
is supported by Deutsche Forschungsgemeinschaft (DFG) under the grant \newline Mu 757/15-1 and by the Belarusian 
Republican Foundation for Fundamental Research under the grant ${\rm \Phi}$07-147. 

\end{document}